\PassOptionsToPackage{numbers,sort}{natbib}
\documentclass[]{ceurart}

\usepackage{listings}
\usepackage{longtable}
\usepackage{array}
\usepackage{hyperref}

\begin{document}

\conference{Preprint. To appear in: PoEM2025: Companion Proceedings of the 18th IFIP Working Conference on the Practice of Enterprise Modeling: PoEM Forum, Doctoral Consortium, Business Case and Tool Forum, Workshops, December 3-5, 2025, Geneva, Switzerland}

\title{Validating API Design Requirements for Interoperability: A Static Analysis Approach Using OpenAPI}

\author[1]{Edwin Sundberg}[%
orcid=0009-0004-2184-9457,
email=edwinsu@dsv.su.se,
]
\address[1]{Stockholm University, Department of Computer and Systems Sciences, Sweden}

\author[1]{Thea Ekmark}[
orcid=0009-0007-2015-4823,
email=thea.ekmark@gmail.com,
]

\author[1]{Workneh Yilma Ayele}[
orcid=0000-0001-8354-4158,
email=workneh@dsv.su.se,
]

\begin{abstract}
    RESTful APIs are central in developing interoperable, modular, and maintainable software systems in enterprises today. Also, it is essential to support system evolution, service interoperability, and governance across organizational boundaries to ensure good quality and consistency of these APIs. However, evaluating API design quality, which is part of non-functional requirement tasks, remains a largely manual and ad hoc process, particularly during early development.
    Using a Design Science Research (DSR) methodology, we elicited user needs, identified 75 API design rules using a literature review, and implemented a configurable rule engine to detect structural violations in OpenAPI specifications. The proposed tool supports organizational adaptability by allowing rules to be customized, enabled, or disabled, enabling integration of domain-specific standards. The evaluation was conducted through structured experiments and thematic analysis involving industry experts.
    API quality validation contributes to aligning technical designs with requirements and enterprise architecture by strengthening interoperability and governance between enterprise systems. The results show that S.E.O.R.A facilitates early validation of non-functional API requirements, provides actionable and traceable feedback, and aligns well with requirements elicitation and quality assurance processes. It improves the API design process by automating checks that would otherwise require manual inspection, thus supporting consistent and reusable conformance practices.
    This work contributes to requirements engineering by operationalizing design principles as verifiable constraints and embedding them into a practical validation tool. Future directions include IDE integration, expanded rule coverage, and real-world deployment to support continuous compliance in agile API development lifecycles.
\end{abstract}


\maketitle

\thispagestyle{empty}

\clearpairofpagestyles
\lehead{\thepage\quad E. Sundberg et al.}
\rohead{Static Analysis for API Design Validation\quad\thepage}

\section{Introduction}
Modern internet infrastructure predominantly comprises web applications supporting enterprise systems such as banking services and other systems, including social media platforms \cite{Serbout22}. These applications are based on web services that expose functionality and data through well-defined application programming interfaces (APIs) \cite{Masse}. Several architectural paradigms exist for designing such APIs, including Service-Oriented Architecture (SOA) and Representational State Transfer (REST), as well as query languages like GraphQL, which offers alternative interaction models \cite{Golmohammadi}. Among these architectural paradigms, REST has emerged as the most widely adopted style since the early 2000s, primarily due to its ability to address scalability concerns and facilitate lightweight development for consumer-facing applications \cite{Tran,Golmohammadi}.

RESTful services are typically documented through human-readable texts or informal diagrams. However, the OpenAPI Specification (OAS), formerly known as Swagger, has emerged as the de facto industry standard for representing RESTful APIs in a machine-readable format \cite{Karlsson,openapiorg,Serbout22}. These specifications define operations, parameters, and responses \cite{Karlsson}, enabling the automated generation of interactive documentation and client libraries \cite{Golmohammadi}.

As REST APIs grow in scale and complexity, their design quality becomes critical to maintainability and adoption \cite{Kotstein}. Developers often choose APIs based on their quality \cite{Masse}. However, as APIs evolve or are updated, violations of best practices or inconsistencies in structure can introduce significant degradation in API quality \cite{Henning,Palma}. Consequently, ensuring that the APIs adhere to design rules is a significant concern for quality assurance, resulting in fewer design issues during maintenance and updates \cite{Kotstein}. Furthermore, existing validation tools, such as Spectral \cite{SpectralLinter}, fail to focus on the underlying structures and design of an API. From an enterprise modeling perspective, RESTful APIs play a critical role in linking business services, processes, and information systems across organizational units and impacts enterprise interoperability. Therefore, integrating API validation into enterprise modeling practices can help ensure alignment between technical specifications and requirements during early design stages.

This paper identifies relevant API Design Rules and proposes a tool named \textbf{S.E.O.R.A} (Static Evaluator of RESTful APIs) to validate RESTful APIs and address the challenges outlined in the previous paragraph. S.E.O.R.A assesses RESTful APIs defined using the OpenAPI Specification against a collection of formalized design rules. The design rules represent quality-related non-functional requirements, which are derived from previous research on REST principles \cite{Palma,Kotstein,Bogner,Palma17,Masse,Tran,Varanasi}. The proposed tool performs static analysis to detect potential design violations, offers actionable recommendations for improvement, and thereby supports the creation of high-quality APIs. Unlike non-static, that is, dynamic evaluation methods, which rely on executing or deployed systems \cite{Arcuri,Palma17},  this approach analyzes API specifications in isolation, making it particularly suitable for early-stage validation and seamless integration into requirements elicitation and quality assurance workflows.

The definition of a ``well-designed'' REST API remains contested \cite{Masse,Tran}. Hence, no agreed-upon definition leads to different interpretations of what constitutes a well-designed REST API \cite{Kotstein}. Prior studies have identified widely accepted design rules and assessed both their significance and the consequences for comprehension when these principles are violated \cite{Masse,Kotstein,Bogner}. These rules can be framed as non-functional requirements that govern how APIs should be structured and evolve.  This paper extends previous efforts by operationalizing these design principles into machine-verifiable rules and embedding them within a practical evaluation framework that supports systematic conformance checking.

\section{Background \& related work}

\subsection{Representational State Transfer (REST)}
\label{rest}
Representational State Transfer (REST), which is a conceptual API framework, defines API behavior \cite{Bogner} using URIs to identify resources and HTTP methods for operations \cite{Golmohammadi,Bogner}. Its simplicity made it popular  \cite{Tran}. REST gained more popularity than Simple Object Access Protocol (SOAP) as service-oriented systems grew \cite{Palma}, enabling more scalable and accessible development \cite{Tran}. A key advantage over SOAP was simplifying web app development by easing data retrieval  \cite{Tran}. The simplicity of REST also supports the implementation of microservices in larger systems \cite{Arcuri}. However, the lack of a formal definition has led to varying interpretations and inconsistent quality of the RESTful API \cite{Kotstein}. 

\subsection{OpenAPI}
\label{openAPI}
The OpenAPI Specification (OAS) defines a RESTful API interface for exchanging information between provider and consumer \cite{Golmohammadi,openapiorg}. It supports two main development approaches: design-first (specification before code, the recommended method) and code-first (specification generated from code) \cite{opeanapiinitiative2}. OAS is language-neutral and allows humans and machines to understand or interpret services without access to source codes \cite{Golmohammadi}. It defines operations, parameters, responses, and data models in JSON or YAML \cite{Karlsson,Golmohammadi,openapiorg}. While alternatives like RAML exist, OpenAPI is the most widely used \cite{Golmohammadi,Karlsson}. The OpenAPI ecosystem supports testing, code generation, and data model analysis\slash validation \cite{Serbout22}. 

\subsection{API evolution} 
\label{backgroundapievolution}
Modern systems, such as microservices, use independently upgradable services, requiring more integration than monolithic architectures \cite{Lercher}. High inter-system dependency complicates API updates\slash modifications \cite{Lercher}, and this added dependency is essential to measure API quality\slash similarity  \cite{Serbout24}. Poor API evolution can degrade quality if bad practices are repeated  \cite{Palma}, making quality evaluation during design and implementation crucial \cite{Kotstein}. However, this evaluation is typically performed manually \cite{Kotstein}. Additionally, domain- and business-specific practices impose extra quality constraints \cite{Tran}.

\subsection{Related research}
\label{relatedresearch}
Previous research on REST API quality includes representing APIs as Java interfaces \cite{Palma}, tested against design principles via static and dynamic checks \cite{Palma}. This approach requires manual definition, and mentioned future work includes expanding its scope and tested principles \cite{Palma}. Also, previous research indicates that there is a significant gap between theoretical and practical API design practices \cite{Rodriguez16}. Furthermore, business requirements\slash policy-driven API design helps ensure consistency and alignment with business requirements \cite{Heshmatisafa23}.

On the other hand, regarding the OAS application, prior work introduced EvoMaster, a tool that generates test cases from an extended OpenAPI version \cite{Arcuri}. Using a evolutionary search approach it achieves high code coverage and detect faults \cite{Arcuri}. Variants of EvoMaster also consider operation dependencies over values and coverage \cite{Karlsson}. For static evaluation, Spectral lints OAS specifications to find issues \cite{SpectralLinter}, focusing on lines rather than API structure or design. This paper focuses on static API evaluation, as prior research mainly emphasizes run-time evaluation or testing. API design rules are hence essential for API evaluations. 

REST API design consistency is supported by several rules based on common patterns in development practices \cite{Masse}. Design inconsistencies can be detected using static and dynamic approaches \cite{Kotstein}. Studies have explored how rule violations impact understandability and relate to core REST principles \cite{Bogner}, offering insight into how inconsistencies affect usability.

\section{Research design}
\label{researchdesign}
The research framework chosen for this paper was Design Science Research (DSR), which focuses on building artifacts to solve real-world practical problems \cite{Perjons}. Artifacts, such as systems, methodologies, and system architectures, developed with DSR allow for an iterative development process using different research strategies and methods for five activities \cite{Perjons}. The five activities in the framework are to explicate the problem, define requirements, design and develop, demonstrate, and evaluate artifacts \cite{Perjons}.

\subsection{Explicate problem}
\label{explicatetheproblem}
A survey in the form of a document review was conducted, which enabled a broad data collection to define the gap between the current and the desired state \cite{Perjons}. This activity aimed to identify important API design rules from academic and gray literature, which was combined with previous research on rule’s priority \cite{Kotstein}.

The performed document study selected seven different academic sources \cite{Kotstein,Bogner,Palma,Palma17,Masse,Tran,Varanasi} and three gray literature sources \cite{Fredrich,InfoQ,StackoverflowJSONNamingConventions}. The document study resulted in the identification of 75 REST API design rules\footnote{\url{https://doi.org/10.5281/zenodo.17447913}}. Each rule received a concise explanation of its meaning and was assigned an identifier. Literature references were provided where the rules were originally documented, and for rules considered untestable, explanations were given for why they could not be tested. When rules were found to be too similar, they were grouped together under a single entry.

We then preliminary categorized the rules after the possibility of implementing detection for them, as ``Yes'' -- if they were deemed detectable and relevant, and ``No'' if not. Common reasons for deeming rules not detectable include requiring run-time access for dynamic analysis capabilities. All rules were then categorized according to their perceived importance as established by \cite{Kotstein}. The distribution of importance levels among the identified rules is presented in \autoref{tab:overviewofidentifiedrulesandtheirranking}.

\newcommand{\calcpercent}[2]{\the\numexpr (#1*100)/#2\relax\%}

\newcommand{\calcrelpercenttotal}[1]{\the\numexpr (#1*100)/75\relax\%}

\begin{longtable}{>{\raggedright\arraybackslash}p{0.22\linewidth}>{\raggedright\arraybackslash}p{0.28\linewidth}>{\raggedright\arraybackslash}p{0.22\linewidth}>{\raggedright\arraybackslash}p{0.18\linewidth}}
    \caption{Distribution of identified rules by implementability and importance level.}
    \label{tab:overviewofidentifiedrulesandtheirranking} \\
    \hline
    \textbf{Implementable} & \textbf{Amount \# of Rules} & \textbf{Importance} & \textbf{\# of Rules} \\
    \hline
    \multirow{3}{=}{Yes} & \multirow{3}{=}{34 (\calcrelpercenttotal{34})} & High & 16 (\calcpercent{16}{34}) \\
    & & Medium & 6 (\calcpercent{6}{34}) \\
    & & Low/N.A. & 12 (\calcpercent{12}{34}) \\
    \hline
    \multirow{3}{=}{No} & \multirow{3}{=}{41 (\calcrelpercenttotal{41})} & High & 11 (\calcpercent{11}{41}) \\
    & & Medium & 9 (\calcpercent{9}{41}) \\
    & & Low/N.A. & 21 (\calcpercent{21}{41}) \\
    \hline
\end{longtable}

\subsection{Define requirements}
\label{definerequirements}
The second activity of the DSR framework was to define the functional and non-functional requirements of the artifact. Semi-structured interviews were conducted with six developers and infrastructure specialists who had professional experience in using and\slash or designing RESTful APIs, see \autoref{tab:respondentsandparticipation}. The respondents represented varying levels of seniority and worked at different companies. Respondent selection employed a purposive sampling based on respondents' relevant expertise in the field.

The interviews were conducted via a digital meeting platform. They included a short introduction about the project, followed by questions on which features the respondents would find helpful when using an artifact such as this. The interviews were analyzed using an inductive thematic analysis approach \cite{BraunClarke}. The interviews were transcribed, coded, and then analyzed to identify themes. The analysis was done on a semantic level and, therefore, based on what was explicitly said by the respondents rather than the meaning \cite{BraunClarke}.

The thematic analysis yielded 82 codes, which were subsequently organized into four main themes: ``Functions'', ``Tool adaptation'', ``Feedback'', and ``Communication of results''.

The theme ``Functions'' included features the respondents mentioned about the artifact's ability to take the ongoing debate about API quality into consideration by being able to customize feature within the artifact. The ability to add, delete, and customize rules was brought up. In the theme ``Tool adaptation'' the respondents mentioned that certain industry-wide and company-wide standards exist. Therefore, having the ability to adjust to company- or industry-specific regulations would benefit the tool's relevance. The respondents mentioned different types of feedback that could be useful when using the artifact; these were included in the theme ``Feedback''. A priority list of warnings to know what is most critical and the ability to make sure that company-wide rules are followed by warning if a rule has been broken were a few of the types of feedback mentioned. The respondents mentioned different ways for the feedback to be visualized, such as CI/CD, command-line, or a dashboard, which were all mentioned under the theme ``Communication of results.''.

\subsection{Design and develop artifact}
\label{designanddevelopartifact}
The third activity of the DSR framework is designing and developing the artifact \cite{Perjons}. There is less of a need for an explicit research strategy for this activity \cite{Perjons}; hence, no explicit strategy is tied to this activity. The artifact was, however, designed and developed iteratively based on the results of the previous activities. The artifact was implemented using technologies previously known and used by us while considering their relevance to the scope of the study. 

\begin{figure}
    \centering
    \includegraphics[width=0.9\linewidth]{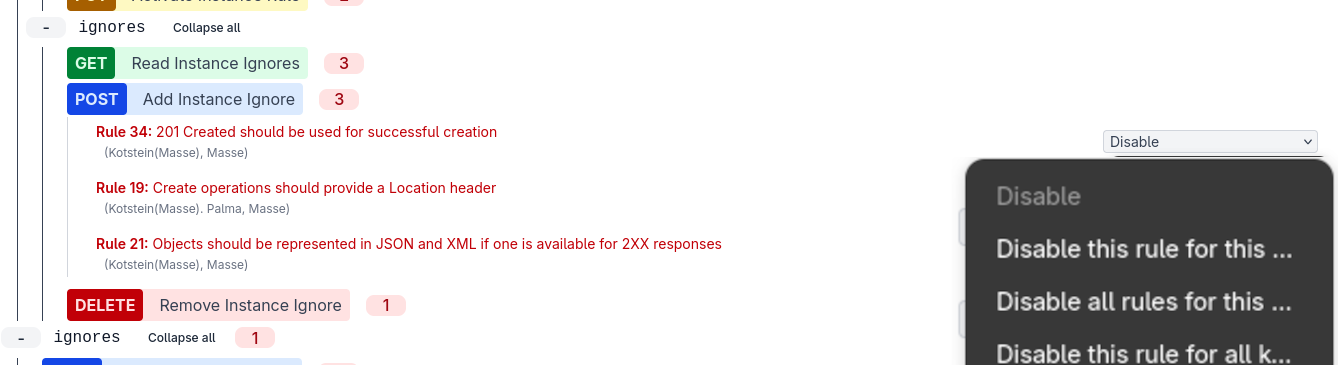}
    \caption{Rule violation interface showing expanded drop-down menu for disabling individual or all rules for a specific key.}
    \label{fig:seora-violation-expanded-dropdown}
\end{figure}

The developed application, S.E.O.R.A \footnote{\url{https://doi.org/10.5281/zenodo.17447848}}, allows users to upload an OpenAPI specification file in JSON \slash YAML format and receive a list of violations and their corresponding rules. A demonstration of the user interface of the application can be seen in \autoref{fig:seora-violation-expanded-dropdown}.

The application consists of a Python FastAPI backend with 23 rule files (covering 34 rules), along with an OpenAPI parser, which is exposed using a RESTful API. Furthermore, a benefit of using Python - i.e., an interpreted language is that more rules could easily be loaded at run-time in future iterations of the artifact. The front-end consists of a React web application styled with Tailwind.

The developed parser followed the OpenAPI version 3.1 specification format \cite{OAS31}, which builds a tree-like structure of the OAS where each node represents a URI key.

\begin{figure}
    \centering
    \includegraphics[width=0.45\linewidth]{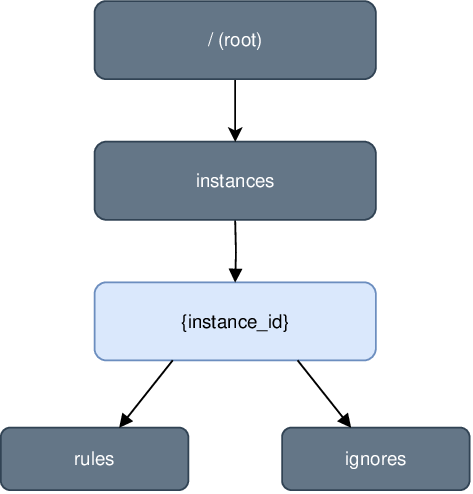}
    \caption{Tree structure visualization of two URIs with highlighted variable \textit{\{instance\_id\}} used across multiple API methods.}
    \label{fig:seora-subset-tree}
 \end{figure}

For example, an OAS with the URIs:
\begin{itemize}
    \item[--] \textit{instances/\{instance\_id\}/rules}
    \item[--] \textit{instances/\{instance\_id\}/ignores} 
\end{itemize}
would be represented as a tree depicted in \autoref{fig:seora-subset-tree}. Additionally, schema and component references are resolved wherever possible to enable more straightforward rule definitions. While resolving schema references substantially increase the parser's output size; this trade-off was deemed acceptable given the simplified rule definitions it facilitated.

\subsection{Demonstrate artifact}
\label{demonstrateartifact}
The fourth activity of demonstrating the artifact was conducted with user tests on three APIs over a digital meeting platform where the respondents controlled a virtual machine with the developed application. In total, eight (8) software developers\slash specialists partook in the user tests, see \autoref{tab:respondentsandparticipation}. The same sampling strategies were used as in \autoref{definerequirements}.

The respondents were given three OpenAPI specifications \cite{fakestoreapi,twilio_sendgrid_mail_api,discord_api_spec} representing varying quality levels, from poor to well-designed APIs per evaluations by the authors and S.E.O.R.A, and asked to input them into the tool and review the corresponding recommendations and warnings generated by the program. The APIs were selected as they provided the users with examples of potential issues in real-life APIs and a fictional case to demonstrate the solution's ability to catch trivial issues. 

Another suitable strategy for this phase would have been the use of a descriptive case study, which would have provided insight into its usefulness in a real-life environment \cite{Perjons}. However, the findings from this type of case study may be influenced by stakeholder perspectives, potentially resulting in biased outcomes \cite{Perjons}. 


\subsection{Evaluate artifact}
\label{evaluateartifact}
The fifth activity of the DSR framework was to evaluate the artifact. An \textit{ex ante} evaluation was used to evaluate this artifact. The framework were evaluated with an inductive thematic analysis that was performed based on the evaluating test and the semi-structured interviews that the eight respondents completed when the artifact was demonstrated. 

The thematic analysis resulted in 32 codes, which were then organized into three themes: ``Improvement opportunities'', ``Integration into workflow'', and ``Appreciated features''. 

During the evaluation test and interviews, the respondents raised some points of improvement under the theme ``Improvement opportunities''. These were abilities such as being able to choose which parts of the API get evaluated and being able to disable violation detection for specific parts of the API. The ability of the artifact to be integrated into users' workflow is important for its relevance; respondents mentioned positive remarks on the artifact's ability to do so but also mentioned some potential improvements to make it even more adaptable. The ability to connect in as a plugin to an IDE or an an integration to the pipeline were a few ideas that were mentioned under the theme ``Integration to workflow''. For the theme ``Appreciated features,'' the respondents mentioned some features that were greatly appreciated, such as the artifact being able to aid a company in checking that they stay consistent and the ability to both modify and add rules. 

\autoref{tab:combinedrequirements} presents the identified requirements from the second activity, ``Define requirements'', along with details regarding their fulfillment obtained during the final activity, ``Evaluate artifact'.

\begin{table}
\caption{Respondents and their participation in the study.}
\label{tab:respondentsandparticipation}
\begin{tabular}{|l|p{0.32\linewidth}|p{0.26\linewidth}|p{0.20\linewidth}|}
    \hline
    \textbf{Resp.} & \textbf{Title} & \textbf{Define requirements} & \textbf{Evaluate artifact} \\
    \hline
    1 & Chief Architect for the Application Area & \textbf{\texttimes} & \textbf{\texttimes} \\
    2 & Infrastructure Specialist & \textbf{\texttimes} & \textbf{\texttimes} \\
    3 & Junior Software Developer & \textbf{\texttimes} & \textbf{\texttimes} \\
    4 & Senior Development Engineer & \textbf{\texttimes} & \textbf{\texttimes} \\
    5 & System Developer & \textbf{\texttimes} & \textbf{\texttimes} \\
    6 & Intern Backend\slash Android Developer & \textbf{\texttimes} & \textbf{\texttimes} \\
    7 & System Developer and Team Lead & & \textbf{\texttimes} \\
    8 & Infrastructure Specialist & & \textbf{\texttimes} \\
    \hline
\end{tabular}
\end{table}

\begin{table}
\caption{Summary of requirements, their fulfillment approach, and the respondents who suggested each requirement.}
\label{tab:combinedrequirements}
\begin{tabular}{|l|p{0.18\linewidth}|p{0.07\linewidth}|p{0.06\linewidth}|p{0.22\linewidth}|p{0.07\linewidth}|p{0.21\linewidth}|}
    \hline
    \textbf{ID} & \textbf{Name} & \textbf{Resp.} & \textbf{Prio.} & \textbf{Description} & \textbf{Impl.} & \textbf{Comment} \\
    \hline
    1 & Detect violations & 1-6 & 1 & Identify design violations using a predefined set of rules. & Yes & \\
    \hline
    2 & Disable rules & 3, 5 & 1 & Provide functionality to deactivate unwanted rules. & Yes & Supports disabling specific rules for individual keys, all rules for a key, or individual rules across the entire API. \\
    \hline
    3 & Enable rules & 3, 5 & 1 & Provide functionality to reactivate previously disabled rules. & Yes & \\
    \hline
    4 & Add rules & 1-2, 4-5 & 1 & Allow addition of custom rules, including organization-specific rules. & Yes & Limited support. \\
    \hline
    5 & Security checks and API specification checks & 1, 3-4 & 3 & Identify and flag potential security vulnerabilities. & No & \\
    \hline
    6 & Generating draft specification & 1 & 3 & Automatically create preliminary API specifications. & No & Not applicable since the underlying API implementation remains unmodified. \\
    \hline
    7 & Configurable rules & 1-6 & 1 & Allow rule customization for context-dependent scenarios. & Yes & \\
    \hline
    8 & Rule violations in priority & 1, 4 & 2 & Display rule violations by severity from most to least critical. & No & Grouping violations by location was determined to be more effective. \\
    \hline
    9 & Warn for potential problems & 5 & 2 & Display alerts for indicators of possible future issues. & Yes & \\
    \hline
    10 & Command-line communication & 2-6 & 1 & Provide results output in command-line compatible format & Yes & Interface implemented, but interaction tools were not developed. \\
    \hline
    11 & Dashboard & 2 & 2 & Create a dashboard to show violations and adjust settings. & Yes & Some RESTful API functionality is missing.  \\
    \hline
\end{tabular}
\end{table}

\section{Discussion}
\label{discussion}
This paper developed a software solution that statically evaluates REST APIs using the OAS format. The aim of this artifact was to improve the evaluation process and, hence, the development process of RESTful APIs. The discussion is structured around the findings of this paper and focuses on the evaluation process and its impact on evolution. Integrating S.E.O.R.A into enterprise modeling environments strengthens requirements elicitation, supports design consistency, and promotes organizational interoperability.

\subsection{API evaluation process}
\label{evalprocess}

The developed artifact performs static evaluation of APIs based on a set of REST design rules. Consequently, it does not require access to a deployed API, as all evaluations are conducted statically on the API specification itself. Therefore, this approach avoids issues related to security and communication between intermediary components \cite{Golmohammadi}. From the evaluation of the artifact, it was found that potential users perceived the software solution as both beneficial and compatible with the typical workflows of developers and specialists working with RESTful APIs. Since assessing API quality is a crucial step that is often carried out manually, the artifact can assist in identifying non-technical design issues early in the development process \cite{Kotstein, Tran}. Furthermore, business- and domain-specific quality constraints can be checked to aid development \cite{Tran}. The artifact can thus facilitate the evaluation of API quality by potentially reducing the manual effort required from developers.

Nevertheless, the current evaluation solution may entail considerable manual effort to disable irrelevant rules and incorporate organization-specific guidelines. Moreover, the rule checks are relatively crude in their detection, which can lead to numerous false positives and interpreting these flagged violations may also demand significant effort to derive actionable insights. Another limitation lies in the user interface, which currently lacks functionality for systematically searching, filtering, or selectively including specific parts of an API specification. Finally, certain design rules, such as the ``Hierarchy design'' compliance level three, are not fully verified due to inherent limitations in the OpenAPI Specification (OAS) format, which does not comprehensively capture relationships between endpoints beyond their hierarchical structure \cite{Serbout22}.

Despite these drawbacks, respondents indicated that the solution could still be valuable if integrated into their routine development workflows. Another benefit of the artifact would be applying it's modularity for policy-driven API design to ensure sustained quality \cite{Heshmatisafa23}.

\subsection{Effect on API maintainability}
\label{apievolution}
The tool provides a method for statically evaluating APIs against a set of established design rules, thereby supporting the maintenance of high-quality API design. APIs assessed by the tool receive a warning for each rule they violate. Each warning includes a descriptive message explaining the violation, which can be interpreted to guide improvements and help prevent poor design quality.

In the context of increasingly modular and interdependent systems, applying an organizational rule set can help ensure standardized and consistent API quality across services. A systematic approach, such as the developed solution, can therefore complement manual reviews by automatically detecting trivial design flaws early, thereby reducing the required effort. However, since the solution evaluates only the API specification, an inherent limitation that remains is that the internal design of the system may still suffer from poor quality even if all design rules applied to the exposed interface are satisfied \cite{Karlsson,Golmohammadi}.

\section{Threats to validity}
\label{threatstovalidity}
To address the threats to \textit{external validity}, respondents for both data collection sessions were developers\slash specialists in API development and usage. The respondents were from different sectors with varying seniority and could therefore provided ifferent perspectives regarding experience and requirements. This variability might have resulted in greater organizational flexibility while potentially sacrificing specialized expertise within individual organizations. In addition, we continued to interview new respondents to collect data until we reached saturation, which occurred when we were unable to find any new information or requirements from subsequent interviews. At this point, we decided to stop, as suggested in \cite{guest2006howmany}. Thus, we determined the number of participants based on the level of saturation. Additionally, although real APIs were used for testing, they were not tested in a real development setting, i.e., \textit{ex-post} evaluation.

While we used ten sources to identify design rules, we argue this provides representative coverage because: (1) we observed convergence in the rules across sources, suggesting theoretical saturation, and (2) these sources (specifically \cite{Masse}) are frequently cited as foundational for design principles in RESTful design quality studies \cite{Kotstein, Bogner, Palma17}, validating their comprehensiveness and authority in the domain

Regarding \textit{internal validity}, i.e., factors that might have affected the results, a controlled environment was used, giving each respondents the same conditions, making the results easily comparable. The controlled setting also ensures the focus of the respondents on study-relevant data while reducing the potential impact of external variables.

Another threat to the internal validity of the study is the process used to decide which rules should be implemented. This was mitigated by including multiple authors when deciding to reduce the risk of the process negatively impacting the results.

\section{Conclusion and future directions}
\label{conclusion}
This study identified a list of API Design Rules from prior research on REST principles \cite{Palma,Kotstein,Bogner,Palma17,Masse,Tran,Varanasi} and translated them into formal, verifiable requirements for RESTful service interfaces. Building on those rules, we developed S.E.O.R.A — a requirements-driven tool that validates OpenAPI specifications before implementation or deployment. By performing static conformance checks, S.E.O.R.A delivers early, actionable feedback that supports requirements elicitation, quality assurance, and enterprise-wide governance of API assets. Software engineers, requirement engineers, and testers could use the proposed tool, S.E.O.R.A, for validating API design quality and enforcing requirements. Our evaluation shows that:
\begin{itemize}
    \item \textbf{Effectiveness for early validation}. Because S.E.O.R.A analyses specifications in isolation, it can be integrated into design reviews, CI pipelines, or requirements workshops to surface rule violations long before code is executed.
    \item \textbf{Adaptability through rule management}. Stakeholders can add, disable, or refine rules, enabling organizations to encode domain- or company-specific guidelines without modifying the underlying engine.
    \item \textbf{Reuse research potential}. The open, tree-based parser and rule framework can be extended in future studies—for example, to compare alternative rule sets, experiment with weighting schemes, or explore other specification formats.
\end{itemize}
At the same time, limitations remain. The expressiveness of the OpenAPI Specification bounds the current approach: relationships between endpoints, behavioral contracts, or cross-service workflows are only partially captured, constraining the complete evaluation of higher-order design principles. Moreover, S.E.O.R.A concentrates on static structure; complementary dynamic analyses (e.g., run-time monitoring or mutation testing) are still necessary for comprehensive quality assurance.

\subsection{Future directions}
\label{future_direction}
Several opportunities exist for further development of the S.E.O.R.A tool and the API Design Rules. Real-world validation of both the tool and its underlying approach are also potential future directions. We suggest the following future directions:
\begin{enumerate}
    \item \textbf{Ex-post evaluation in real-world projects} While this study used controlled experiments and ex-ante evaluation to assess S.E.O.R.A's utility, future work should include ex-post evaluation in live software projects. Embedding the tool in actual API development workflows (e.g., during a sprint or release cycle) would offer insights into long-term adoption, usability, rule effectiveness, and the impact on defect prevention. Observational studies and longitudinal assessments could measure real-world outcomes such as improved design consistency, reduced review time, or higher developer confidence. This could be conducted as an A/B test to compare speed and accuracy.
    \item \textbf{Integration and deployment with other tools} S.E.O.R.A could be integrated into CI/CD pipelines using tools such as GitHub and GitLab for automated enforcement of design conformance. It is also possible to package S.E.O.R.A as an Integration Development Environment (IDE) plugin to provide inline feedback similar to Github co-pilot. S.E.O.R.A could also be implemented as a web service and provisioned as a Software-as-a-service offering to support centralized API governance functions and distributed teams. This future implementation not only improves the accessibility of the tool but also enables organizational-level rule standardization and reuse.
    \item \textbf{Rule engine enhancement and coverage expansion} The current rules focus primarily on structural and naming conventions. Future work could expand the rule base by including security-related rules, versioning policies, etc. Also, it is possible to introduce rule prioritization schemes based on stakeholder feedback or other relevant parameter.
    \item \textbf{Integration into model-driven and enterprise application development tool} Positioning S.E.O.R.A within a model-driven engineering (MDE) or enterprise architecture (EA) environment (e.g., ArchiMate, UML) could bridge requirements elicitation with conformance validation. Violations could be visualized directly within service models, aiding traceability and compliance reporting.
    \item \textbf{Explore how S.E.O.R.A can be embedded into enterprise modeling workflows} Linking API-level validation to higher-level business process models.
    \item \textbf{Cross-specification generalization} While S.E.O.R.A targets OpenAPI 3.1, many modern ecosystems include diverse interface formats. Future iterations could:
    \begin{itemize} 
    \item Investigate interoperability rules across heterogeneous interfaces to support federated or event-driven systems.
    \item Add support for GraphQL schemas, enabling multi-spec validation under a unified framework.
    \end{itemize} 

\end{enumerate}

\section*{Declaration on Generative AI}
During the preparation of this work, the author(s) used ChatGPT, Claude and Grammarly in order to: Grammar and spelling check, Paraphrase and reword. After using these tool(s)/service(s), the author(s) reviewed and edited the content as needed and take(s) full responsibility for the publication’s content. 

\bibliography{bibtex}

\end{document}